\documentclass{aastex}
\usepackage{emulateapj5,apjfonts,epsfig}

\def\asca{{\sl ASCA }}

\def\xte{{\sl RXTE }}

\def\chandra{{\sl Chandra }}

\def\cirx1{Cir~X-1}
\def\ergsec{\hbox{erg s$^{-1}$ }}
\def\ergcm{\hbox{erg cm$^{-2}$ s$^{-1}$ }}

\def\kms{\hbox{km s$^{-1}$}}
\def\fexxv{Fe~{\sc xxv}}
\def\fexxvi{Fe~{\sc xxvi}}
\def\mgxi{Mg~{\sc xi}}

\def\Msun{$M_{\odot}$ }

\def\mdot{${\dot M}$ }
\def\it{\sl}

\def\lapp{\ifmmode\stackrel{<}{_{\sim}}\else$\stackrel{<}{_{\sim}}$\fi}

\def\gapp{\ifmmode\stackrel{>}{_{\sim}}\else$\stackrel{>}{_{\sim}}$\fi}
\def\spose#1{\hbox to 0pt{#1\hss}}

\def\approxlt{\mathrel{\spose{\lower 3pt\hbox{$\sim$}}
        \raise 2.0pt\hbox{$<$}}}
\def\approxgt{\mathrel{\spose{\lower 3pt\hbox{$\sim$}}
        \raise 2.0pt\hbox{$>$}}}

\slugcomment{Submitted for publication to The Astrophysical Journal}

\shorttitle{X-RAY LINE EMISSIONS CIR~X-1}

\shortauthors{SCHULZ ET AL.}

\begin{document}

\title{The Origins of X-ray Line Emissions in Circinus~X-1 at Very Low X-ray Flux}
\author{
N. S. Schulz\altaffilmark{1},
T. E. Kallman\altaffilmark{2},
S. Heinz\altaffilmark{3},
P. Sell\altaffilmark{4},
P. Jonker\altaffilmark{5},
and
W. N. Brandt\altaffilmark{6}}
\altaffiltext{1}{Kavli Institute for Astrophysics and Space Research, 
Massachusetts Institute of Technology,
Cambridge, MA 02139.}
\altaffiltext{2}{Goddard Space Flight Center, NASA,
Greenbelt, MD .}
\altaffiltext{3}{Department of Astronomy, University of Wisconsin,
Madison, WI .}
\altaffiltext{4}{Department of Physics, University of Crete, Heraklion, Greece}
\altaffiltext{5}{Department of Astrophysics, Radboud University Nijmegen, The Netherlands} 
\altaffiltext{6}{Department of Astronomy \& Astrophysics, 525 Davey Laboratory,
The Pennsylvania State University, University Park, PA, 16802.}

\begin{abstract}
Accretion conditions and morphologies of X-ray transients containing neutron stars 
are still poorly understood. Circinus X-1 is an enigmatic case where 
we observe X-ray flux changes covering four orders of magnitude. We observed
Circinus X-1 several times at its very lowest X-ray flux using the high energy
transmission grating spectrometer on board the Chandra X-ray Observatory. At a
flux of 1.8$\times10^{-11}$ \ergcm we observed a single 1.6 keV blackbody spectrum.
The observed continuum luminosity of 10$^{35}$ \ergsec is about 
two orders of magnitude too low to explain the observed photoionized luminosity suggesting
a much more complex structure of the X-ray source which is partially or entirely obscured as
had been previously suggested. This affects most emissions from the accretion disk
including previously observed accretion disk coronal line emissions.
Instead, the strongest observed photoionized lines are blueshifted by about
$\sim 400$ \kms\ and we suggest that they originate in the ionized wind of a B5Ia supergaint
companion supporting a previous identification. The neutron star in Cir X-1
is very young and should have a high magnetic field. At the observed luminosity
the emission radius of the blackbody is small enough to be associated
with the accretion hot spot as the X-ray emitting region. 
The small emission radius then points to a field strength below $10^{12}$ G which would be
consistent with the observation of occasional type I X-ray bursts at
high magnetic fields. We discuss Cir X-1 in the
context of being a high-mass X-ray binary with some emphasis on a possible 
Be-star X-ray binary nature.
\end{abstract}


\section{Introduction}

Since its discovery in the early days of X-ray astronomy ~\citep{margon71},
Cir X-1 has shown a vast range of brightness levels, variablity patterns,
and spectral changes in its X-ray emissions. Despite significant advances in recent years
these emissions remain poorly understood. Until about a decade ago it appeared
fairly well established
from photometric variability of the optical counterpart~\citep{stewart91, glass94},
its orbital parameters~\citep{brandt95, tauris99}, as a well as its 
X-ray spectral and timing patterns ~\citep{tennant87, shirey99} that Cir X-1 is 
probably a low-mass X-ray binary containing a neutron star. The latter
is now well determined through the direct observation of type I X-ray
bursts~\citep{tennant86} and their recent confirmation ~\citep{linares10, papitto10}.  
From kinematic parameters and the 
assumption that Cir X-1 is associated with the supernova remnant G321.9-0.3,
~\citet{tauris99} deduced a companion mass of about 2 \Msun or less and a
very extreme orbital eccentricity (e $\sim$ 0.9). More recently~\citet{jonker07} 
determined that the companion is more massive, most likely an A0 to B5 type
supergiant and revised the orbital eccentricity to a more moderate value
(e $\sim$ 0.45) reviving an orginal identification by \citet{whelan77}.

The picture definitely changed when \citet{heinz13} revealed a faint 
X-ray supernova remnant associated with Cir~X-1. This  
allowed the determination of the age of the system to about 4500 years, 
which has two important consequences besides making Cir X-1 the 
youngest X-ray binary of its class known today. It implies
that the neutron star should have a magnetic field exceeding 10$^{12}$
Gauss~\citep{kaspi10}, which is at odds with the occasional observation
of type I X-ray bursts in this source. Such events are ususally observed
in old low-mass X-ray binaries with magnetic fields below 10$^{9}$ Gauss
and where the accretion stream is hardly affected by such a low field.  
Such youth also re-affirms the determination that the companion is massive because
a low mass star could not have had time to evolve to fill its Roche-lobe
at periastron. In a core collapse, the progenitor star was also massive~\citep{jonker07},
the companion star should then be a massive main sequence star which is 
a few tens of Myrs old. Once the binary orbit is largely circularized  
Cir X-1 will eventually become a high-mass X-ray binary as we know them today.
The possibility that the supernova was caused by an accretion
induced collapse (AIC,~\citet{bhattacharya91}) is unlikely as
neutron stars in electron capture supernovae are not expected to
receive a significant kick and is inconsistent with the dynamic orbital
parameters and evolution of the system \citep{clarkson04, tauris13, heinz13}. 

In X-rays, Cir X-1 exhibits two main variability patterns,
one happens on short time scales related to its orbital period, another one 
spans over many years with X-ray fluxes changing from mCrab
levels to several Crab. One orbit lasts about 16.5 days~\citep{kaluzienski76} and
for most of its orbital tenure, X-ray emissions are fairly persistent. Due to 
its orbital eccentricity, the neutron star and its accretion disk actually
spend most of the time detached from the companion star. Near zero phase 
(periastron) the Roche-lobe of the companion overflows and the neutron
star/disk system attaches to the overflow stream and actively accretes matter.
This results in a significant rise in X-rays at periastron passage, which at times
can radiate up to super-Eddington fluxes~\citep{brandt96, brandt00, schulz02}. 
The frequency and 
strengths of these periastron flares also seem to follow a long-term variability
pattern that spans over 30 years~\citep{parkinson03}.   

This long-term lightcurve reflects the flux changes of 30 years measured
with most of the major detectors and observatories available up to 2001. In the
mid-1990s Cir X-1 was as bright as 2 Crab and exhibited relativistic radio jets
~\citep{stewart93, fender98}. \asca observations showed that the accretion disk
is probably viewed fairly edge-on~\citep{brandt96}. One of the biggest revelations
was delivered by the first \chandra observations, which showed strong P Cygni lines
indicating a powerful and variable accretion disk wind ~\citep{brandt00, schulz02}.
~\citet{heinz07} discovered a parsec scale X-ray jet (see also \citet{sell10}) 
manifesting the picture that
at times of high flux Cir X-1 behaves like a true micro-quasar~\citep{mirabel01}.
Since then the X-ray source has dimmed steadily. In 2005 the source was already
down in flux by more than an order of magnitude. While the P Cygni lines were gone,
an emission line spectrum emerged rich in H- and He-like lines from high-Z elements
such as Si, S, Ar, Ca, and Fe at very high neutral columns. There was a notable 
presence of an ionized absorber as well as some resonant absorptions indicating 
a rather weak disk wind. Since these observations the X-ray flux continued
to drop into levels below 10 mCrab. In a rare event the X-ray source at periastron
experienced a larger outburst in 2010 during which Cir X-1 exhibited type I X-ray bursts
~\citep{linares10, papitto10}. This period was extensively covered with \xte
and one \chandra observation providing much temporal and spectral information
but no definite conclusions with respect to the origins of the X-ray continuum
~\citep{dai12}. However it confirmed that Cir X-1
is a neutron star X-ray binary. 

In this paper we analyze a series of observations between 2008 and 2017 which
were taken at the absolute lowest flux levels in decades with the goal
to characterize the nature of the X-ray source at these levels as well as
to find interfaces to the new evolving picture that Cir X-1 is a very young massive
main sequence X-ray binary.

\section{Chandra Observations}

\cirx1 was observed with the high energy transmission grating spectrometer
(HETGS, see \citet{canizares05} for a detailed description)  
once in 2008 and a few times in 2017. Table~1 summarizes all its parameters. 
The 2008 observation we denominate as 'V', the 2017 observations as 'VIIa-c'.
We add one more observation done in 2010 (PI: D'Ai) and label this
one as 'VI'. This is in line with the denominations defined in ~\citet{schulz08}
in sequence to the previous \chandra HETGS observations. Observations I and
II are described in ~\citet{brandt00} and \citet{schulz02}. observations
III and IV in ~\citet{schulz08}.
Observations V and VIIa-c were all performed during periastron passage, observation
VI was done at apastron passage.

\begin{table*}[t]
\begin{center}
{\sc TABLE~1 CHANDRA X-RAY OBSERVATIONS}
\begin{tabular}{llccccc}
            & & & & & & \\
\tableline
\tableline
 \chandra &  Obs. & Start Date & Start Time & Exposure & Phase & HETG 1st rate    \\
  ObsID   &  ID  &    (UT)        & (UT)       & (ks)     & ...    & (cts s$^{-1}$)   \\
\tableline
          & & & & & & \\
  8993 & V    & July 16 2008 & 07:57:55 &  32.7 & periastron & 0.12 \\
 12235 & VI   & July 04 2010 & 05:04:05 &  19.4 & apastron   & 1.13 \\ 
 18990 & VIIa & June 06 2017 & 12:35:21 &  13.2 & periastron & 0.14 \\
 20093 & VIIb & June 06 2017 & 22:00:54 &   9.8 & periastron & 0.15 \\
 20094 & VIIc & July 07 2017 & 06:50:07 &  35.3 & periastron & 0.22 \\
         & & & & & & \\
\tableline
\end{tabular}
\end{center}

\end{table*}

\vspace{0.3cm}
\includegraphics[angle=0,width=8.5cm]{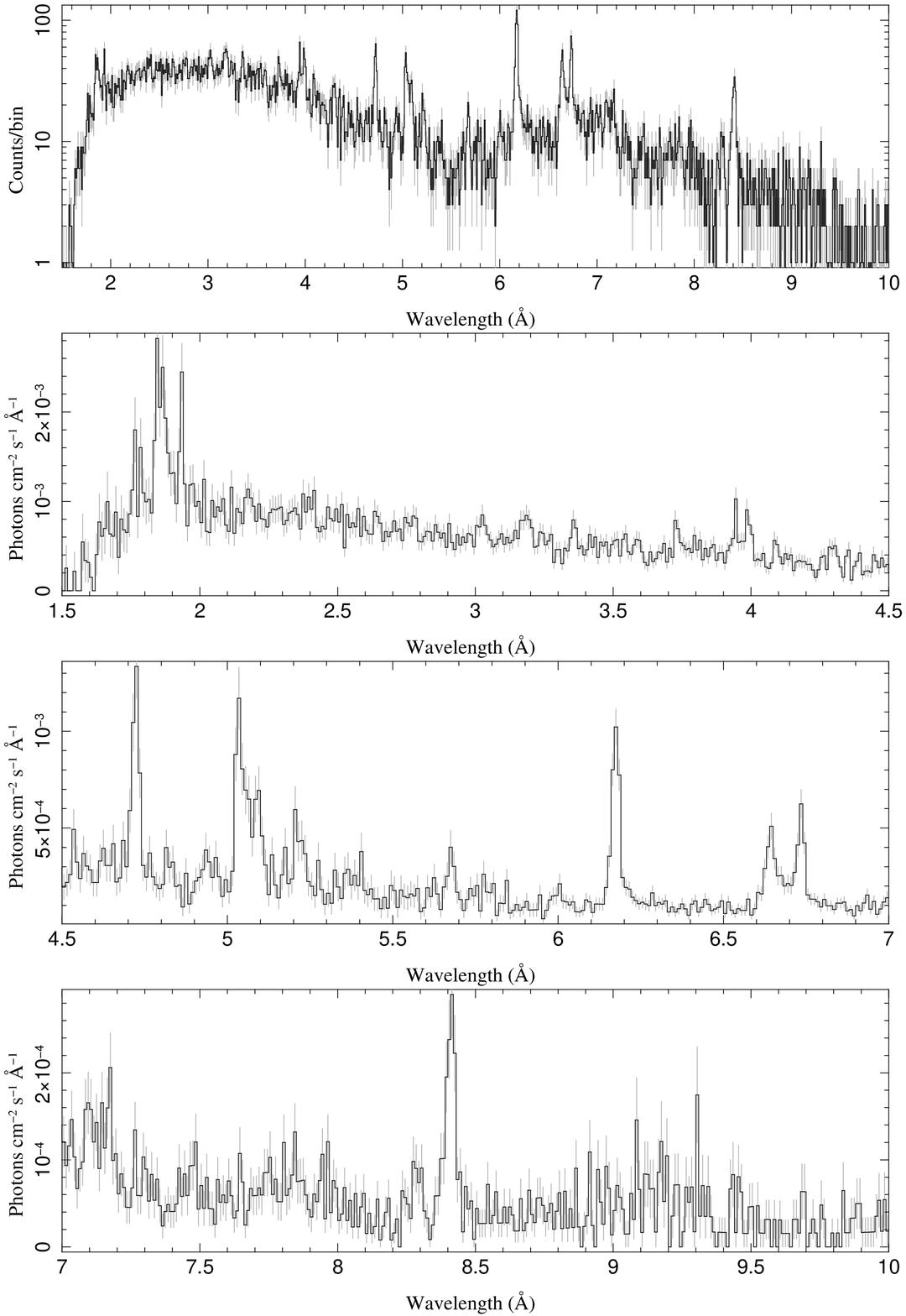}
\figcaption{
The raw-count HETG 1st order spectrum of the co-added observations (top panel).
A more spread out unfolded spectrum of the HETG 1st order (bottom three panels). 
V and VIIa-c.
\label{figure1}}
\vspace{0.3cm}

All observations were reprocessed using CIAO 4.9 using CIAO
CALDBv4.7.7 products. Updates to CIAO and its products since then did not
impact the analysis preformed at the time of submission. 
The wavelength scale was determined by measuring the zero-order
position to a positional accuracy of about half a detector pixel ensuring a wavelength
scale accuracy of about a quarter resolution element, i.e. 0.005~\AA\ for MEG and 
0.003~\AA\ for HEG spectra. For our previous observations we
could not use a direct zero-order source detection because the point spread function of the 
source was too piled up. In that case we measured the zero position by
determining the intersection between the readout streak and the grating dispersion arms.
In this analysis we applied both methods and their results agreed well within 
the accuracy stated above. Note, in observation III
~\citep{schulz08, iaria08} we did not even have a source readout streak which 
introduced a systematic uncertainty that resulted in slight redshifts in the 
detected lines. It is thus paramount that we maintain a high confidence in the 
location of the zero order position.
For transmission gratings the dispersion scale is linear in
wavelength, therefore we perform all analysis in wavelength space. This ensures
the most accurate scales through multiple binnings.
We used standard wavelength redistribution matrix files (RMF)
but generated ancillerary response files (ARFs) using the 
provided aspect solutions, bad pixel maps, and CCD window filters.
\footnote{see \url{http://asc.harvard.edu/ciao/threads/}}
For all the observations we
generated spectra and analysis products for the medium energy gratings (MEG) +1 and -1
orders, as well as for the high energy gratings (HEG) +1 and -1 orders. Figure~\ref{figure1}
shows co-added 1st order HETG spectra for all periastron observations
(V and VIIa-c) binned by a factor 4, which represents about one
MEG resolution element per spectral bin. The 
lightcurves of the periastron observations are all very similar and flat without any structure and
we therefore do not present any plots. Observation V is very faint with 0.12 c/s, VIIa,b are
similar, 
observation VIIc is slightly brighter (see Table~1). The apastron observation (VI) was
taken during a specifically strong outburst and is an order of magnitude
brighter than all the other observations. This observation is shown in Figure~\ref{figure7}.

\vspace{0.3cm}
\includegraphics[angle=0,width=8.5cm]{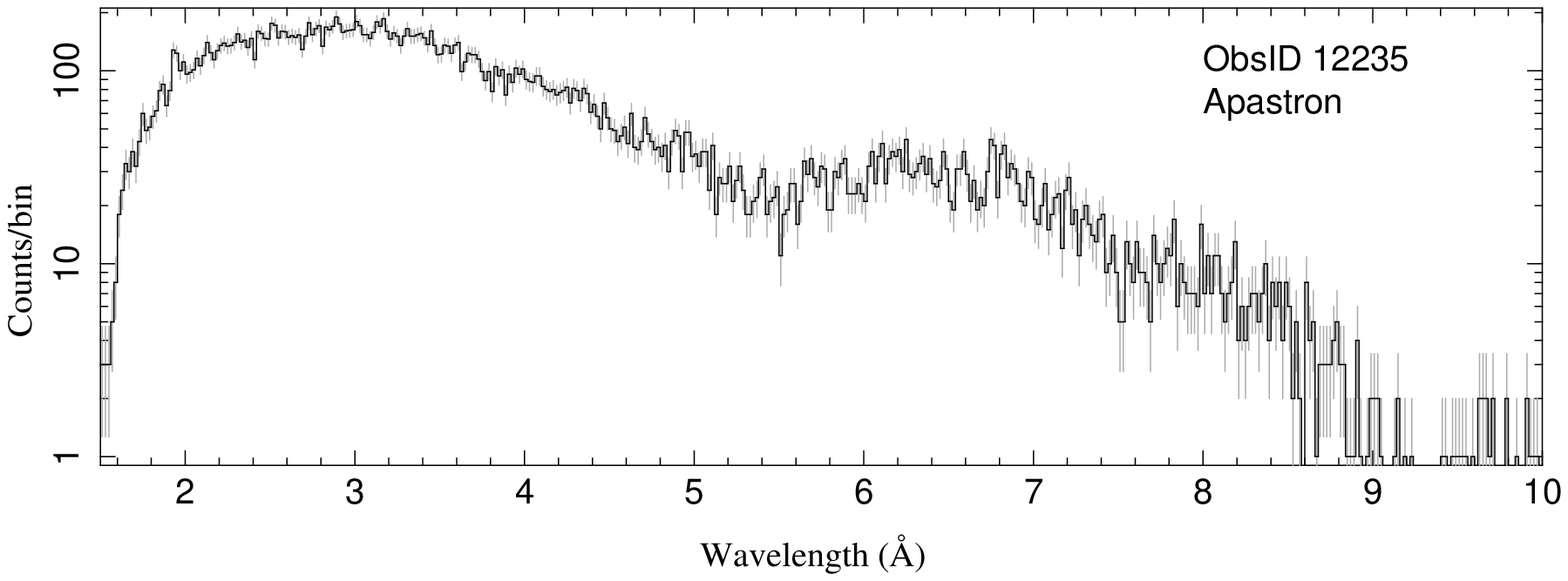}
\figcaption{The X-ray spectrum at apastron during an intermediate outburst in 2010.
\label{figure7}}
\vspace{0.3cm}

\section{Spectral Analysis}

For the spectral analysis we focus exclusively on the four periastron observations.
Each individual observation does not have enough statistics to perform detailed
line analysis and in general we co-add all of them in the plotting and fit 
simultaneously in the analysis. Due to the fact that these
observations were over an order of magnitude fainter than the ones in 
~\citet{schulz08} we used a \emph{Cash} statistic~\citep{cash79} for the fits.  
This only reflects a more accurate treatment of low statistics data while
still allowing for a consistent comparison with previous much brighter
observations.


Figure~\ref{figure1} shows faint and bright lines in the spectrum. For the
determination of the continuum we remove these wavelength regions from the 
spectral model fit. The observed lines all appear at known line wavelengths
for Mg, Si, S, Ar, Ca and Fe and they also appear quite narrow. We removed
0.05 \AA\ regions around the detected line centroids for the 
line centroids (see Sect.~3.2) detected for these elements. 
All spectra then have 1011 data bins in the unbinned case. We do not bin the 
spectra during the fits as this introduces statistical biases. However, we do
match the grids of the HEG to the MEG spectra to plot the combined spectrum.

\vspace{0.3cm}
\includegraphics[angle=0,width=8.5cm]{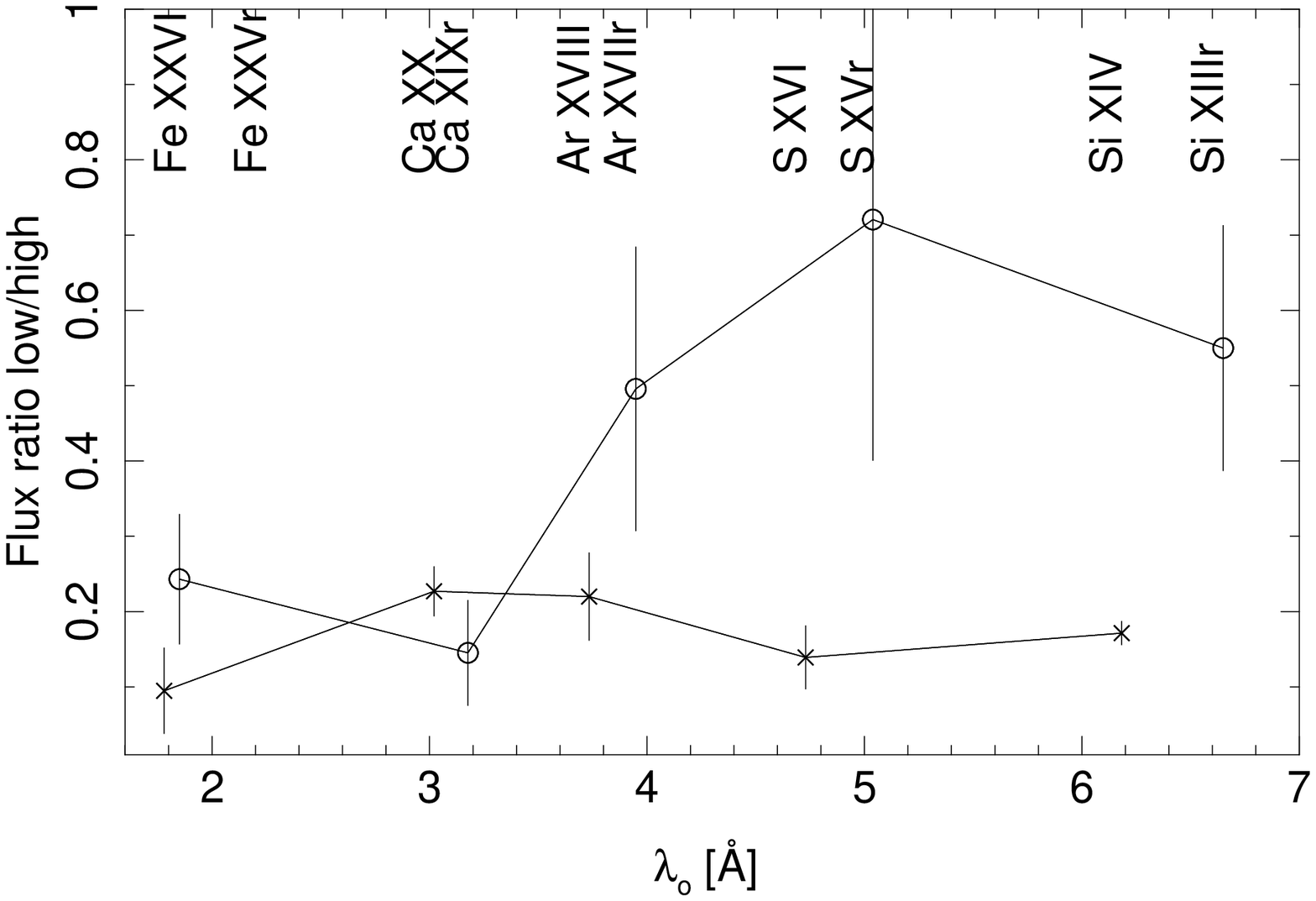}
\figcaption{Line ratios of the H- and He-like resonance lines
from the high flux observations in \cite{schulz08} and the low
flux observations reported in this paper. 
\label{figure2}}
\vspace{0.3cm}


\subsection{X-ray Line Analysis}

The co-added spectrum in Fig~\ref{figure1} shows very strong H- and He-like lines
for Mg, Si, and S with line counts between 150 and 320 cts/line. This is large enough
to allow the
determination of critical line parameters such as line shifts, widths, and flux.
For this analysis we use Gaussian line functions. The 
H-like lines and the Fe K line are single Gaussians. We do not account 
for the spin-orbit split in the H-like lines which produces a separation of 
the $\alpha_1$ and the $\alpha_2$ line by 0.0056~\AA\ and use the 
average line location as was done in ~\citet{schulz02} and ~\citet{schulz08}. This
allows us to better compare the line measurements.
As a consequence we may slightly overestimate the line width in H-like lines.
The He-like lines are triplets containing a resonance (r), an intercombination (i),
and a forbidden (f) line with fixed line spacings. Here we froze the line 
spacings of the i and f lines relative to the r lines. 
We then divided the bandpass into five regions containing the Mg, Si, S,
Ca + Ar, and Fe lines, respectively. 

\begin{table*}[t]
\begin{center}
{\sc TABLE~3: X-RAY LINE PROPERTIES FROM THE PERIASTRON OBSERVATIONS IN TABLE 1} \\
\begin{tabular}{lcccc}
  & & & &\\
\tableline
\tableline
 ion & $\lambda_o^(1)$ & $\lambda_{maes}$  & Line Flux                       & $\sigma$ \\ 
     & \AA         & \AA               & 10$^{-5}$ ph s$^{-1}$~cm$^{-2}$ & \kms \\ 
\tableline
  & & & &\\
Fe~XXVI           &   1.780 &   1.782$\pm$0.010 & 0.92$\pm$0.99 & $<$3300\\
Fe~XXVr           &   1.850 &   1.845$\pm$0.008 & 3.84$\pm$2.23 & 1000$\pm$820\\
Fe~XXVi           &   1.859 &   1.858$\pm$0.008 & 0.10$\pm$8,38 & 1000$\pm$820\\
Fe~XXVf           &   1.868 &   1.870$\pm$0.005 & 3.25$\pm$3.60 & 1000$\pm$820\\
Fe~K              &   1.940 &   1.933$\pm$0.001 & 2.49$\pm$0.85 & 390$\pm$430\\
Ca~XX~L$\alpha$   &   3.021 &   3.021$\pm$0.008 & 0.59$\pm$0.44 & 770$\pm$640\\
Ca~XIXr           &   3.177 &   3.171$\pm$0.007 & 0.30$\pm$0.36 & 750$\pm$510\\
Ca~XIXi           &   3.189 &   3.183$\pm$0.007 & 0.74$\pm$0.55 & 750$\pm$510\\
Ca~XIXf           &   3.211 &   3.205$\pm$0.007 & $<$ 0.1       & 750$\pm$510\\ 
Ar~XVIII~L$\alpha$&   3.734 &   3.736$\pm$0.003 & 0.88$\pm$0.35 & 500$\pm$460\\
Ar~XVIIr          &   3.949 &   3.946$\pm$0.004 & 1.19$\pm$0.39 & 600$\pm$310\\
Ar~XVIIi          &   3.966 &   3.963$\pm$0.004 & 0.14$\pm$0.50 & 600$\pm$310\\
Ar~XVIIf          &   3.994 &   3.991$\pm$0.004 & 1.32$\pm$0.49 & 600$\pm$310\\
S~XVI~L$\alpha$   &   4.730 &   4.722$\pm$0.002 & 2.95$\pm$0.55 & 450$\pm$140\\
S~XVr             &   5.039 &   5.036$\pm$0.003 & 2.45$\pm$0.54 & 470$\pm$130\\
S~XVi             &   5.067 &   5.064$\pm$0.006 & 0.97$\pm$0.44 & 470$\pm$130\\
S~XVf             &   5.102 &   5.093$\pm$0.003 & 1.32$\pm$0.43 & 470$\pm$130\\
Si~XIV~L$\alpha$  &   6.183 &   6.175$\pm$0.001 & 2.64$\pm$0.28 & 420$\pm$70\\
Si~XIIIr          &   6.650 &   6.640$\pm$0.003 & 1.32$\pm$0.24 & 500$\pm$80\\
Si~XIIIi          &   6.669 &   6.678$\pm$0.003 & 0.63$\pm$0.34 & 500$\pm$80\\
Si~XIIIf          &   6.740 &   6.734$\pm$0.002 & 1.40$\pm$0.24 & 500$\pm$80\\
Mg~XII~L$\alpha$  &   8.422 &   8.409$\pm$0.004 & 0.94$\pm$0.18 & 530$\pm$140\\
Mg~XI(r)          &   9.169 &   9.179$\pm$0.015 & 0.25$\pm$0.14 & $<$ 480\\
Mg~XI(i)          &   9.230 &   9.228$\pm$0.012 &  $<$ 0.1      & $<$ 480\\
Mg~XI(f)          &   9.314 &   9.315$\pm$0.013 & 0.22$\pm$0.15 & $<$ 480\\
  & & & &\\
\tableline
  & & & &\\
(1) see \citet{schulz08}
\end{tabular}
\end{center}

\end{table*}

\subsubsection{Line Fluxes}

The results of the line fits to the periastron data in Table~1 are 
shown in Table~3. The first columns list the K-shell ion, the second column
the theoretical line location as in \citet{schulz08}, 
the third column is the measured wavelengths, 
the fourth column gives line fluxes in units of $10^{-5}$ photons s$^{-1}$ cm$^{-2}$,
the fifth column lists the sigma of line widths in \kms 
(3$\times10^5 (\sigma_{meas}/\lambda_{meas}$)). There are distinct differences
in the line fluxes with respect to previous observations when the source was brighter.
Figure~\ref{figure2} shows that the bulk of the H-like line fluxes was 
of the order of 20$\%$ of the ones reported in \citet{schulz08}, while the
He-like lines are up to 70$\%$ of the previously reported fluxes. Calcium lines are
the exception as they were generally weak in observations III and IV \citep{schulz08}. 
The lines in the Fe K line region follow this picture, the H-like \fexxvi\ is very weak
and the region is dominated by the \fexxv\ triplet. However, the Fe K line (Fe~I - X)
is very similar in flux and width to the one we observed in observation III.
The line widths cluster around 550 \kms, which is very similar 
to previous detections. The line widths at shorter wavelengths appear
higher which is likely a consequence of the increasingly lower spectral 
resolution at shorter wavelengths. The size of an HETG
resolution element of a dispersed grating order is constant in wavelength
space and thus resolution element sizes increase in velocity towards higher energies.

\vspace{0.3cm}
\includegraphics[angle=0,width=8.5cm]{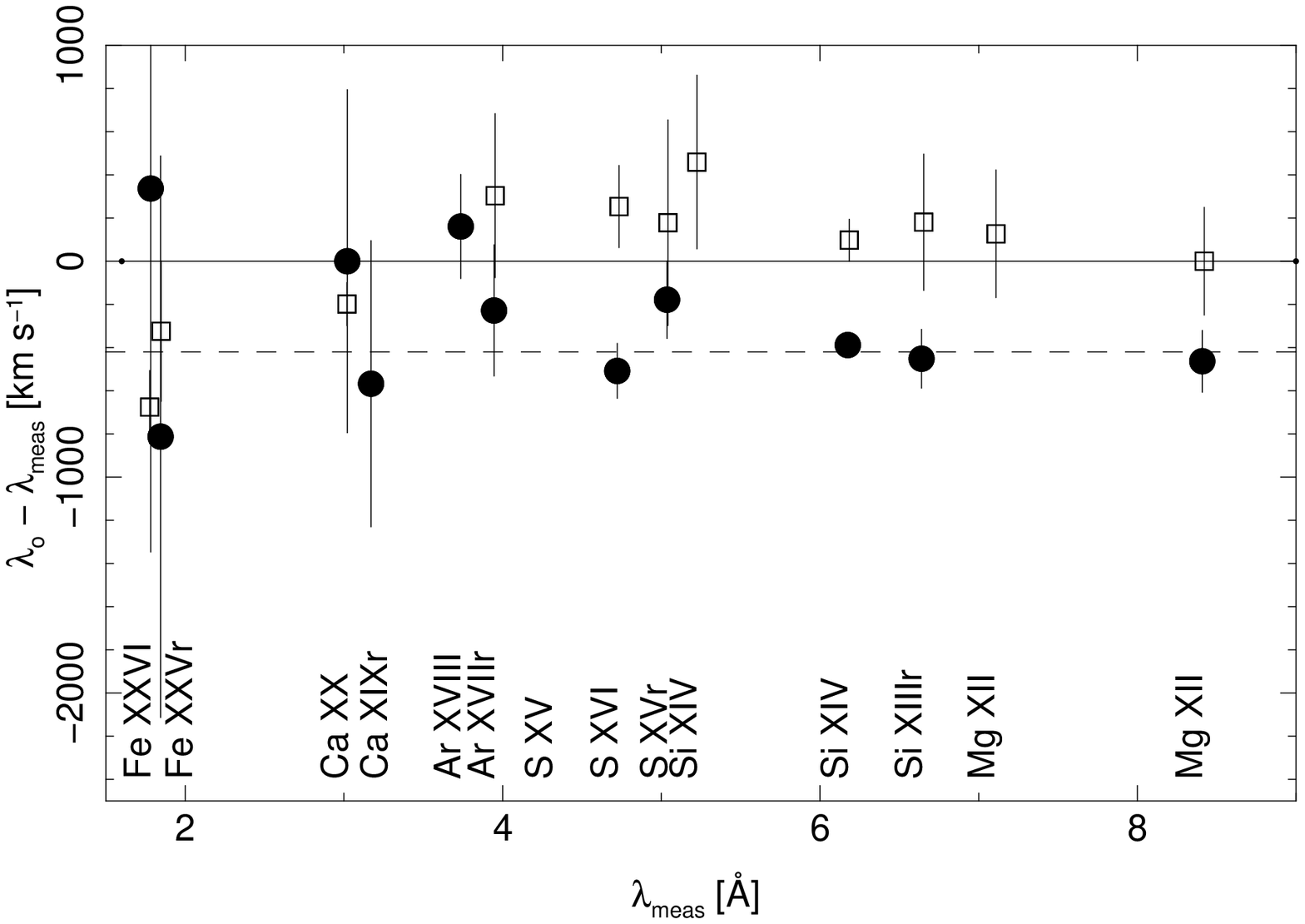}
\figcaption{Measured line centroids with respect to theoretical values
in velocity units. Negative values are blueshifts. The straight line
are zero velocities from the theoretical line values. The dashed line
marks the average blueshift of -410 \kms seen in the periastron observations
V, VIa-c. The squares are the data from ~\citet{schulz08} for observation
IV for reference. 
\label{figure3}}
\vspace{0.3cm}

\subsubsection{Line Centroids}

Figure~\ref{figure3} shows the line centroids with the
expected rest positions in units of \kms. The filled black circles show
the periastron observations from Table~1. The majority of the brightest lines
concerning Mg, Si and S clearly show a line shift to the blue of about 400 \kms.
Higher Z element lines (Ar, Ca, and Fe) are fainter and uncertainties are larger. 

For comparison we plot the results from observation
IV from ~\citet{schulz08} for reference (squares). 
For this observation we had to refit the He-like 
triplets (Fe, Ca, Ar, S, and Si) because in that 
previous analysis the widths and line spacings were free parameters and not 
tied according to their triplet properties. In some cases, specifically for
the Si, Ar, and Ca  triplets, this produced line centroids that are much more consistent
with the rest of the sample. The line centroids from observation IV are 
consistent with the expected rest wavelengths.  

\vspace{0.3cm}
\includegraphics[angle=0,width=8.5cm]{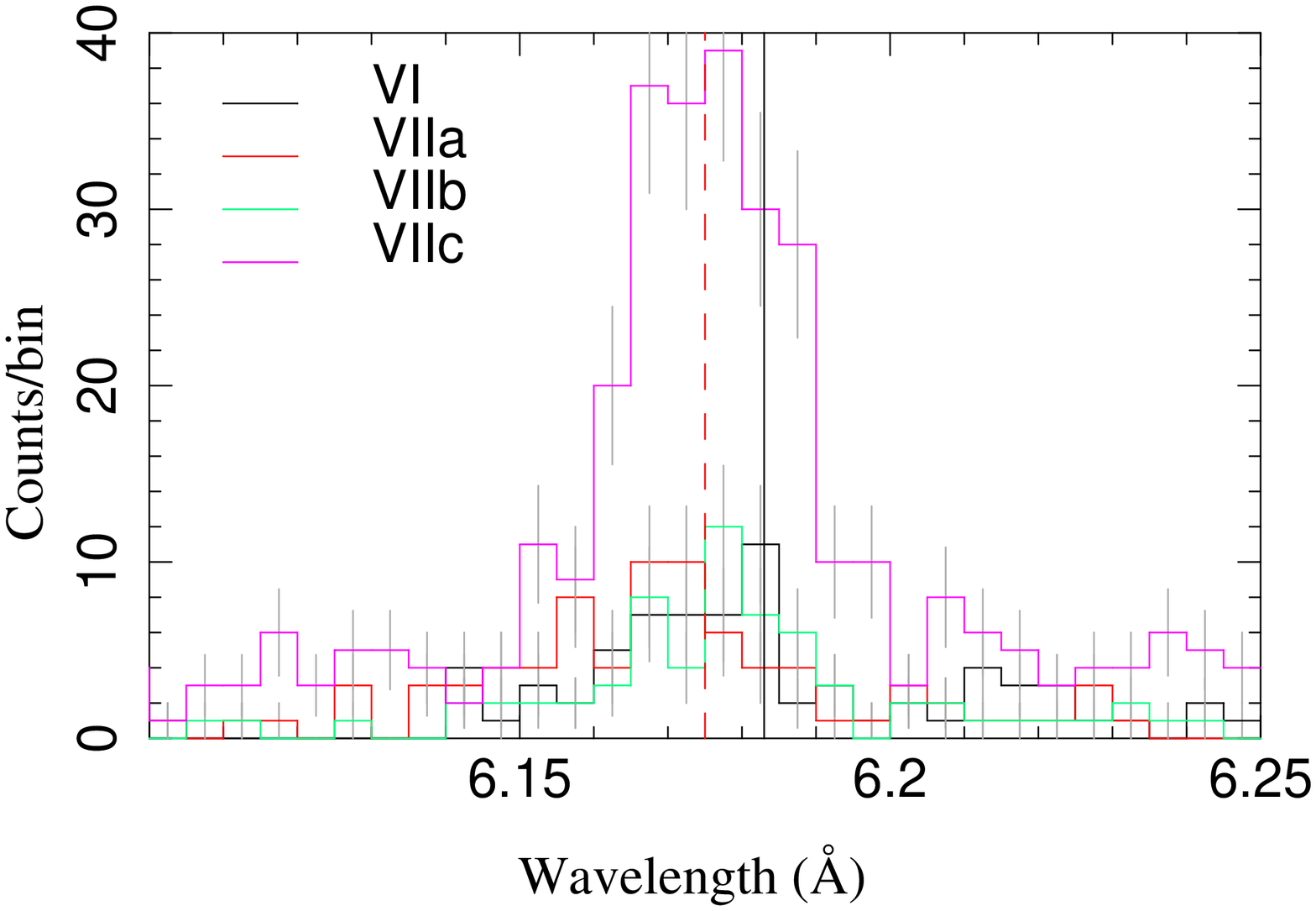}
\figcaption{The Si~XIV L$_{\alpha}$ line for all four observations
plotted separately. The hatched line marks the centroid location of the 
co-added line fits. The solid line marks the rest wavelength location.
It shows that all observations agree with a line centroid location 
blue-shifted by about 400 \kms.
\label{figure4}}
\vspace{0.3cm}

The Si~XIV L$_{\alpha}$ at the expected wavelengths is by far the most 
significant one in terms of total counts ($>$ 300 cts) and here we can look
at the indivdual observations. The brightest contribution comes from observation
VIIc ($\sim$ 170 cts), the faintest from observation V ($\sim$ 35 cts), enough
to allow centroid studies. Figure~\ref{figure4} shows the line counts of the 
unbinned data. The black line marks the expected rest wavelength at 6.183~\AA, 
the hatched line the fitted location of the co-added line data. It clearly shows 
that the blue-shift is present in all observation segments and 
appears persistent over many years. The lines of observations V and VIIa,b are symmetric
around the fitted centroid (hatched line). 

\subsubsection{G and R Ratios}

Line flux ratios are a powerful tool to diagnose detailed properties of
an ionized gas~\citep{bautista00, porquet00}. 
A gas at temperatures of $< 1$ MK that is photoionized is expected to emit lines
primarily by recombination cascades and will have a G-line flux ratios of
$G = (f+i)/r > 4$ or close to that value for atomic numbers $Z$ of
12 and higher. From the fluxes in Table~3 we   
find G values of 1.53$\pm$0.30, 0.93$\pm$0.60, 1.22$\pm$0.49, 2.46$\pm$1.11, 0.87$\pm$1.07 
for Si, S, Ar, Ca, and Fe, respectively. These values appear well below the 
expectation for a pure photoionized plasma indicating some form of hybrid 
plasma. We rule out EUV photoexcitation as this would depopulate
the $\alpha$ resonance line and produce higher order transitions which do
not appear in Table~3. There are two processes that can lead to enhanced 
resonance line emissions, one is collisionally ionization in very hot plasmas,
the other one is resonance scattering.
We can rule out contributions from a collisionally ionized plasma
as here the excitation of K-shell ions with atomic number larger than
14 would also require temperatures in excess of 10 MK and the presence of 
a significant bremstrahlungs continuum which we do not observe. 
What we observe is likely photoionized plasma affected by resonance scattering.
In that case the presence of a high optical depth medium can scatter
resonance line fluxes unisotropically into preferential directions depending
on geometry. Modeling of this effect requires extensive knowledge
of the geometry and local plasma properties.   

Gas densities can be diagnosed with the
$R = f/i$ ratio for which we expect that at $ R > 1$ densities are below
a critical density for a specific atomic number $Z$. Our sample features
He-like triplets for Si, S, Ar, Ca, and Fe and features critical desities
of $\sim10^{14}, \sim10^{14}, \sim10^{15}, \sim10^{16}$, and $\sim 10^{17}$ cm$^{-3}$.
In all cases we find R values much larger than 1 except for Ca, where we only
detect an upper limit for the f line. 
This points to plasma densities lower 
than $\sim5 \times 10^{14}$ cm$^{-3}$ assuming an isotropically mixed gas.  
The R ratio is also very susceptible to UV radiation where the metastable
forbidden line levels get depopulated into intercombination line levels. Since
the f-lines in Table~3 generally appear stronger than the corresponding i-lines
shows that effects from UV radiation appear not to be significant. 

The lines in the Fe K line region follow this picture, the H-like \fexxvi\ is very weak
and the region is dominated by the \fexxv\ triplet. However, the Fe K line (Fe~I - X) 
is very similar in flux and width to the one we observed in observation III.

\subsection{Photoionization Modeling}

\begin{table*}[t]
\begin{center}
{\sc TABLE 4  THE PHOTOIONIZATION FIT RESULTS} \\
\begin{tabular}{lccc}
 & & & \\
\tableline
\tableline
Parameter & Symbol & Units &  \\
\tableline
Absorption column density & $N_{H}$   & 10$^{22}$ cm$^{-2}$ & 1.8  \\
Blackbody temperature     & $kT$      &   keV            & 1.57$\pm$0.04 \\
Blackbody normalization   & $(R_{km}/D_{10kpc})^2$ & ...   & 0.392$\pm$0.026 \\
Pcf absorption column 1   & $N_{H}^{1}$  & 10$^{22}$ cm$^{-2}$ & 2.4$\pm$0.4 \\
Photoionization normalization 1  & $K_1$  & erg cm$^{-1}$ s$^{-1}$  & 121.5$\pm$12.7  \\  
Ionization parameter 1    & $\xi_1$ & erg cm s$^{-1}$ & 340$\pm$135\\
Redshift 1                & $z_1$    &  ... & -0.0013334  \\
Pcf absorption column 2   & $N_{H}^{2}$  & 10$^{22}$ cm$^{-2}$ & 2.8$\pm$1.1 \\
Photoionization normalization 2  & $K_2$  & erg cm$^{-1}$ s$^{-1}$  & 100$\pm$65 \\              
Ionization parameter 2    & $\xi_2$ & erg cm s$^{-1}$ & 1500$\pm$180  \\
Redshift 2                & $z_2$    &  ... & -0.00062 \\
Fit statistic             & Cash/dof & ... &  1.42  \\
            & & &\\
\tableline
\end{tabular}
\end{center}

\end{table*}

Figure~\ref{figure1} features 
strong He-like lines with especially strong r-lines. The ionization parameter is defined
as

\begin{equation}
\xi = L_x/(n d^2),
\end{equation}

\noindent
where L$_x$ is the source luminosity, n the plasma density, and d the distance
from the X-ray source. A multiple plasma environment requires a range
of ionization parameters to engage in a fit. This complicates the modelling process 
in terms of fitting time and parameter range.
In that light we try to limit the number of fit components to a very
minimum. For the choice of continuum we modeled a few cases including combinations
of powerlaws as reported by \citet{schulz08} and \cite{iaria08} as well as
blackbody spectra. 
For the fit itself we use \emph{XSTAR}'s \emph{photemis} function
available in \emph{Xspec}, which allows for a pre-set atomic levels population file
to calculate a photo-ionized spectrum for a single ionization parameter. For the final fit 
setup we used two different functions to account for two ionization parameters.
Furthermore in order to be able to fit H- and He-like line morphology, the functions
also needed individual absorption columns, for which we applied \emph{pcfabs} functions. 
For the fits we fixed the interstellar column to 1.8$\times10^{22}$ cm$^{-2}$ as reported
by \cite{heinz13}. The fits generally converged to a one component solution for the 
contiuum, i.e. either a powerlaw or a blackbody spectrum. The powerlaw solution was
ruled out because it completely overshot the observed continuum above 2 \AA\ for the 
Fe K line region, whereas the 1.6 keV blackbody provided the necessary steep decline
above 2 \AA\ (see Fig.~\ref{figure10}). The fit result is summarized in Table~4.   
The uncertainties are 90$\%$ confidence limits calculated by \emph{conf$\_$loop} in
\emph{ISIS}. In the fit we also set the turbulent velocities to 600 \kms\ and
in the final stages we fixed some of the abundances we let float. The redshifts were also
fixed to -400 \kms (see dashed line in Figure~\ref{figure3}) in the one of the 
\emph{photemis} components, let afloat in the second. The fact that we need at least
two ionization parameters is due to the fact that we observe Mg and Si lines as
well as Ca and Fe lines, which cannot be modeled with a single ionization
parameter~\citep{kallman04}. The Mg abundance was
slightly reduced to 0.93, which likely is a consequence of the statistically
challenging \mgxi\ bandbass. The Si abundance was set to 1.73, the S abundance to 1.68
for both \emph{photemis} components.  

\vspace{0.3cm}
\includegraphics[angle=0,width=8.5cm]{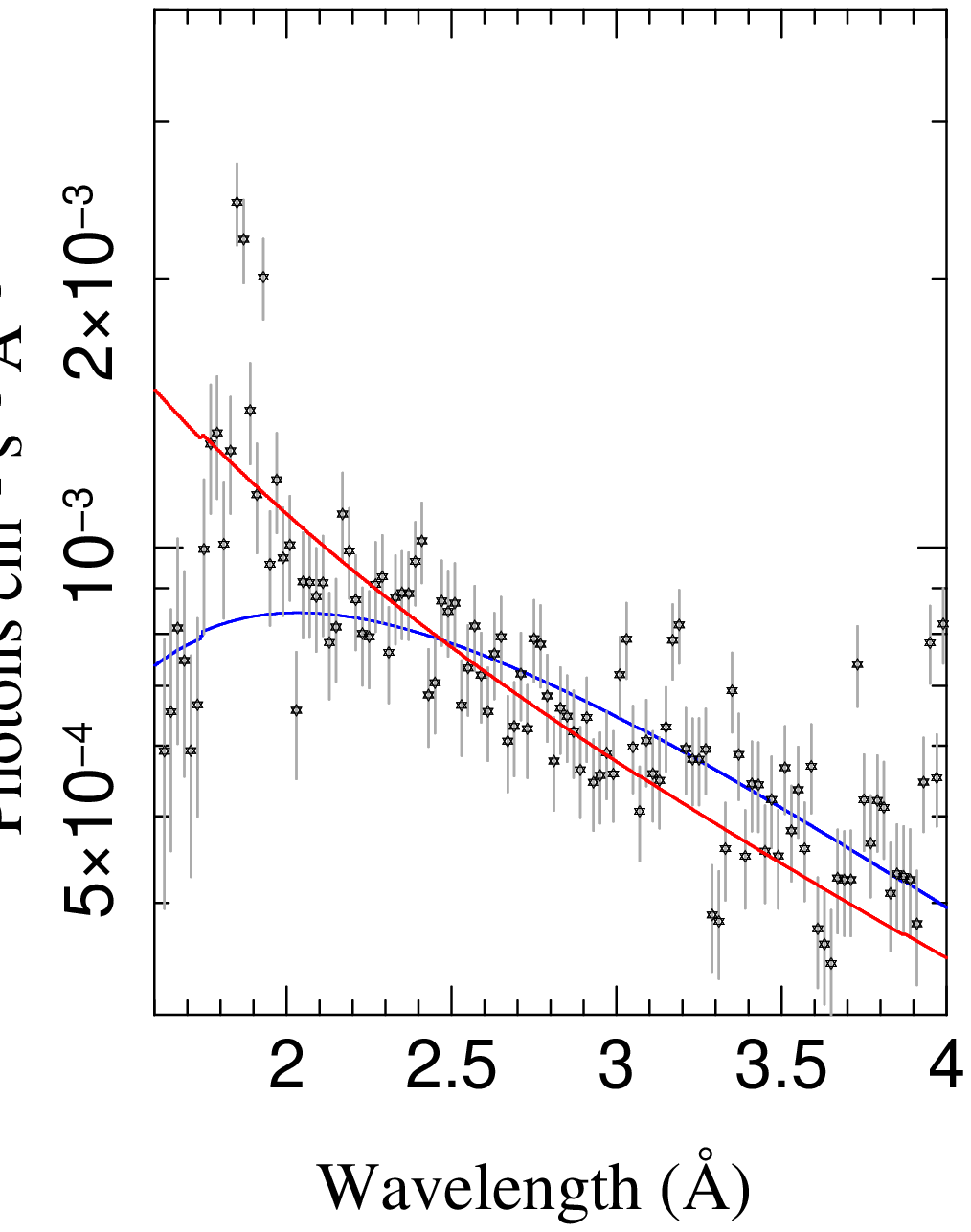}
\figcaption{
Comparison of the powerlaw fit (red) to the blackbody fit (blue) in the
Fe K line region (the data are (black)).
\label{figure10}}
\vspace{0.3cm}

The final ionization parameters obtained were $\xi$ = 340 and 1500 erg cm s$^{-1}$ 
and both associated with 
high column densities. The covering fractions were free parameters but resulted in 0.99 in
both cases. The plasma with ionization parameter 340 erg cm s$^{-1}$
is responsible for bulk of the 
line fluxes, while the one with ionization parameter 1500 erg cm s$^{-1}$
contributes to some
of the H-like line fluxes but mostly to the Fe region.

\vspace{0.3cm}
\includegraphics[angle=-90,width=8.5cm]{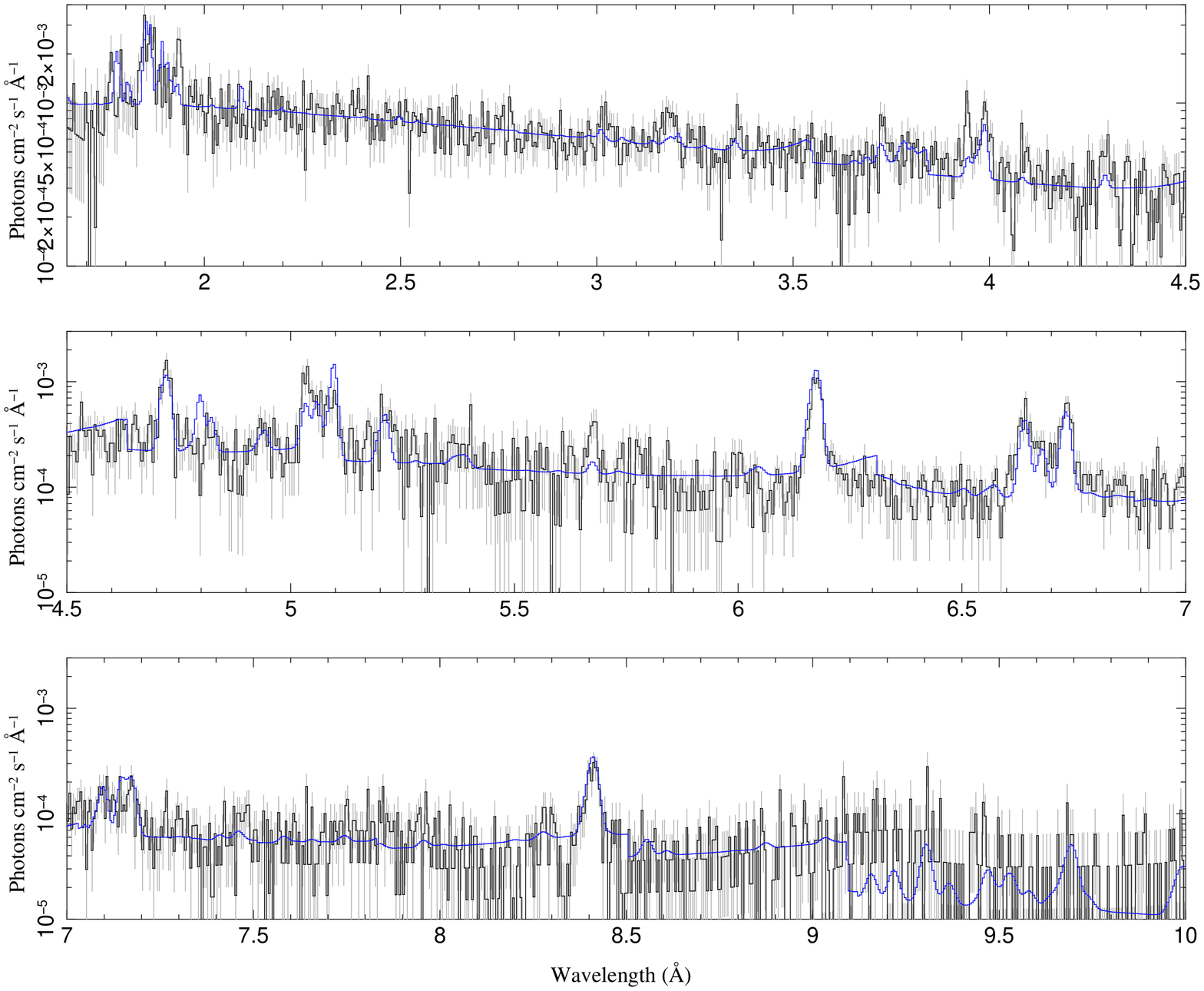}
\figcaption{Photo-ionization model fit using \emph{XSTAR:photemis} function.
\label{figure6}}
\vspace{0.3cm}

The normalization of the photoionization model is defined by

\begin{equation}
K = {{EM} \over 4\pi D^2}\times10^{-10},
\end{equation}

\noindent
where EM is the emission measure of the gas in units of cm$^{-3}$ at
the involved ionization parameter, D is the distance to Cir X-1 in units
of cm. For a distance of 9.4 kpc~\citep{heinz15} this leads to

\begin{equation}
EM = K \times 1.06\times10^{56}.
\end{equation}

\noindent
For the plasma with log $\xi_1$ = 340 erg cm s$^{-1}$ we obtain 
EM$_1 = 1.3\times10^{58}$ cm$^{-3}$,
for log $\xi_2$ = 1500 erg cm s$^{-1}$ 
we obtain EM$_2 = 1.1\times10^{57}$ cm$^{-3}$. 
These emission measures are only slightly smaller than the one estimated
by ~\citet{schulz08}, the observed source luminosity, however, appears 
to be significantly lower.

This observed low X-ray luminosity from the continuum
is inconsistent with the observed photo-ionized luminosity from
the X-ray line analysis. 
One of the difficulties is to maintain the observed 
ionization parameters at such a low luminosity over such a large volume.
While to some extent one could offset a larger source distance $d$ with
a lower density in the wind $n$, this eventually breaks down due to the
fact that the observed photo-ionized luminosity appears close to the actual
source luminosity. Such a high ionization efficiency is nearly impossible
to achieve under likely any circumstances. This points into the direction that 
parts of the X-ray source is still obscured, something we encountered during
the analysis of observations I and II ~\citep{schulz08}. To simply 
assume that some the flux of our observed continuum is partially blocked 
and should be higher is possible but not a good solution
as a 1.6 keV blackbody is inheritently inefficient
to ionize high Z atoms such as Fe and Ca at any possible luminosity.  
A more promising scenario relates back to our actual findings in ~\citet{schulz08}. 
There we found that the observed photoionized lines are due to an 
accretion disk corona (ADC) and in order to sustain such a corona the
source X-ray luminosity needed to be significantly higher than observed
which lead to the conclusion that parts of the central X-ray source 
are obscured. This may still be the case. Figure~\ref{figure9} shows a 
recalculation of the content of Figure 5 in ~\citet{schulz08}, including 
lower luminosity predictions and the locations of our photo-ionized
data points in this picture. This indicates that for our results that the
source still is at a luminosity
close to 10$^{37}$ \ergsec, which allows to sustain a atatic ADC.
This then leaves the possibility that we do observe some highly
absorbed and obscured parts of the ADC in the high Z lines and
the photo-ionized wind of the companion seen in the blue-shifted
lower-Z lines.

The ionization parameter and emission measures then leave us with  
a wide range of viable plasma regimes. Two regimes that are opposite in character
are of interest in the case of Cir X-1. One refers to the ADC emissions as described in
\citet{schulz08} which are confined to the domain of the accretion disk. 
In that case we can assume that {\it d}, the distance
to the X-ray source, is of the same order of magnitude as {\it r}, the 
size scale of the photo-emitting region. For distances of up to 10$^{10}$ to 
10$^{11}$ cm from the neutron star surface we obtain ADC plasma densities
between about 10$^{12}$ and 10$^{14}$ g/cm$^{3}$ in a compact volume 
within the accretion disk (see also ~\citet{jimenez02}).

The observed blueshifts point to another regime.
For very low densities as they exist in weak stellar winds,
i.e. densities $<< 10^{11}$ g/cm$^{3}$ we also get viable solutions
for values of {\it d} and {\it r} that are beyond the dimensions
of the accretion disk. Weak low density winds are expected in mid-B type stars. 
For such low plasma densities the dimensional parameters are
larger than $10^{12}$ cm consistent with ionized line emissions
in a weak stellar wind. 

\vspace{0.3cm}
\includegraphics[angle=90,width=8.5cm]{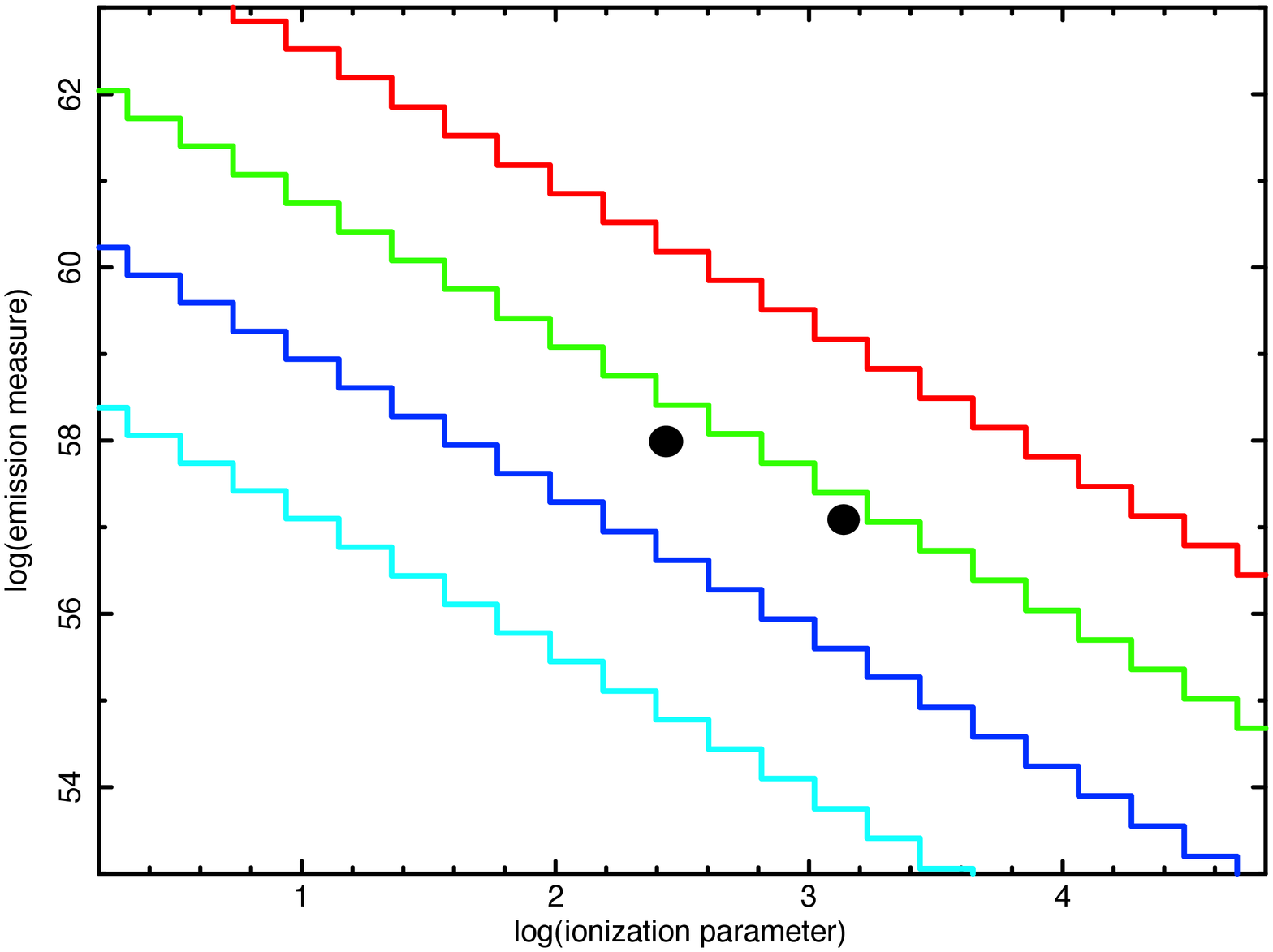}
\figcaption{Calculation of emission measure in depedence of ionization
parameter for a static corona as was done in ~\citep{schulz08}. The
color lines are for various luminosities, $log L_x$ = 38 (red), 37 (green),
36 (blue), and 35 (light blue) in units of \ergsec. The two black 
circles are the results from the photo-ionization modeling.
\label{figure9}}
\vspace{0.3cm}
 
The \emph{XSTAR:photemis} function applies a model describing a purely photo-ionized 
plasma. The fact that this model cannot properly account for the strength of the resonance
lines in some of the He-like triplets confirms our
findings from the G-ratios that there is a significant contribution from some recombination
and resonance scattering. Future modeling of these effects should reveal the presence of
such a scattering medium,

\section{Discussion}

The X-ray source in the Cir X-1 binary has given us many different
looks in the past in terms of variability, flux, and spectral properties.
Since the first \chandra HETG observation in the year 2000, where the
source flux still exceeded 1.5 Crab, the X-ray source has dimmed
by more than three orders of magnitude until it went below one mCrab in
2009 (see Figure~3 in \citet{heinz13}). Its spectral continuum shape similarly
has morphed from two partially covered blackbodies during its high state
\citep{brandt96, schulz02}, to two or one 
partially covered powerlaws~\citep{schulz08, iaria08}, and now 
to a single blackbody with just interstellar
absorption. Besides exhibiting various
levels of line emission and absorption there always was the presence of
enormous levels of either partially covered cold or warm intrinsic 
continuum absorption. High levels of source intrinisic absorption are also present in
the obervations we report in this paper,
except in these new observations these absorbers have only  
little effect on the continuum and mostly affect the photoionized regions. 
This is an aspect we have not seen 
before and it appears to be fundamentally different from the previous
observations. 
 
\subsection{The Origin of the X-ray Continuum in this Very Low State}

The final continuum spectrum in the photo-ionization fit turned out to 
be a 1.6 keV blackbody. We had fixed the interstellar column to 
1.8$\times10^{22}$ cm$^{-2}$ which was found by ~\citet{heinz13} 
in the fit to the supernova remnant, consistent with the large amount of
visible extinction towards Cir X-1 as well as a well established distance
of 9.4 kpc~\citep{heinz15}.
Such a distance is also consistent with Cir X-1 radiating at Eddington peak fluxes when
it was brightest~\citep{jonker04} and when it exhibited P Cygni X-ray lines from a radiation
driven wind~\citep{brandt00}. The property of the blackbody fit is rather
peculiar as it exhibits a very small emission radius. It is also notable
that there is no or only very little contribution of the accretion disk 
to the observed X-ray spectrum. The only clear signature from the disk may come from
the Fe K fluorescence line observed at 1.93 \AA.

The big problem arises when we realize that our observed continuum
is quite insufficient to photo-ionize the plasma at the level observed. 
This gives reasons to assume that emissions from the accretion disk are
there but are highly absorbed and obscured. This provides quite some
uncertainty to the observed luminosity and nature of the X-ray spectral
continuum as we have to consider that the bulk of the emission partly blocked and obscured.  

Accreting neutron stars with magnetic fields significantly lower than 
$10^{10}$ Gauss are generally seen in LMXB, which are considered to be 
older systems in which the original field had enough time to decay to such
low values. With only a very few exceptions where we do not know the
field through cyclotron lines accreting neutron stars with high mass
companions have all high magnetic fields as they are considered very young.
However, we point out that the question of the
companion nature is fairly irrelavant here.
\citet{homan10} showed that transient sources
can morph through atoll and Z-stages at various flux phases making these
spectral variability imprints more related to effective Roche-lobe overflow
accretion rather than binary types. In the following we discuss how the 
observed continuum rates to cases of low and high magnetic fields for the
accreting neutron star.

\subsubsection{The Low Field Case}
 
One of the reasons why Cir X-1 used to be considered a LMXB was its spectral
variation pattern during its brightest flux phases showing the nature of a
Z-source. We then may compare it to a rare class of LMXB pulsars such as
the transient Terzan 5 (IGR J17480-2446) or the persistant ultracompact
binary pulsar 4U1626-67. At an estimated magnetic field between 10$^9$ and 10$^{10}$ G
Terzan 5 is much older and 
its accretion stream is likely only weakly affected by the magnetic field.
As a consequence its spectral signatures are expected to be more comparable to neutron star
atmospheres or normal LMXB emissions depending on whether the transient is
in a subcritical or critical state (see \citet{degenaar13} and references
therein). 4U 1626-67 with a magnetic field of 4$\times10^{12}$ G ~\citep{orlandini98}
is much younger but since its luminosity is persistently close to critical
its continuum emission relates more to what we observed in Cir X-1 during
its intermediate flux phases~\citep{schulz08}. 

A significantly higher blackbody luminosity also significantly 
increases the emission radius to sizes that are more reminiscent of
neutron stars with low magnetic fields in which the accretion flow is hardly
affected by the magnetic field and we can identify random hot patches or
boundary layers on the neutron star surface as emission regions.
Such blackbody continua are not particularly unusual in LMXB, prime examples are
4U1626-67 \citep{schulz01}, 4U1822-37 \citep{ji11}, Her X-1 \citep{ji09} or 
Aqu X-1 \citep{sakurai14}.
However, in all of these cases that blackbody component
exhibits temperatures of about 0.5 keV. High temperatures like observed
here are more prominent in the peaks of type I X-ray bursts and in fact the 
here observed continuum temperature is close to what \citep{linares10} report 
for the top temperatures observed in the 2010 X-ray bursts from Cir X-1.
That is very unusual for LMXBs. The fact that this continuum is heavily 
absorbed and some of this absorption in unaccounted for changes little,
as it is the part $< 2$ \AA\ that defines the blackbody in the fit and
its temperature. We note that neither \citet{schulz08} nor
\citet{iaria08} observed a blackbody when the source was an order of
magnitude brighter but report a more suitable powerlaw for the observed
photoionizations as the continuum. 
The shape of a 1.6 keV blackbody is not particularly suited for K-shell
ionizations of high Z atoms as its shape drops dramatically above 7 keV and
lacks the harder photons.
Should the neutron star have a low field
as is common in LMXBs then the observed continuum properties are quite
unusual and hard to explain, it would also make Cir X-1 the youngest
neutron star with a low magnetic field known to date.

\subsubsection{The High Field Case}

However, what is very common is the fact that young neutron stars, accreting or
isolated, have high ($>> 10^{10}$ G) magnetic fields \citep{reig11, oezel16}.
In this case we consider the possibility that this observed blackbody
continuum does not contribute much to the photo-ionization process at all
and the photo-ionizing continuum is entirely obscured.
In a high field accretion scenario matter is predominantely funneled
by the field onto poles of the neutron star and the state of the
accretion column depends on X-ray luminosity (see \citet{becker07} for 
and indepth review). At a critical luminosity of L$_{crit} = 1.5\times10^{37}$ \ergsec
the accretion flow becomes radiation pressure dominated and X-ray emissions
are defined by a complex mix of physical processes resulting in
something close to a power law with an exponential cut-off or similar as
observed when the source was brighter~\citep{schulz08}. However, in the
sub-critical case where the luminosity is $<< 10^{37}$ \ergsec as is observed
in the low state of Cir X-1, the emissions from the bare hot spot can
be explained by such a hard blackbody.
 
In such a sub-critical environment
we can expect that the X-rays are produced by impact onto or very close to the 
neutron star surface. 
If this is the accretion hot spot, its radius r$_0$ should fulfil the condition~\citep{lamb73}

\begin{equation}
r_0 < r_{ns} \lbrack{r_{ns}\over r_A}\rbrack^{1/2},
\end{equation}
    
\noindent
where r$_A$ is the Alfven radius given by~\citet{becker07} as

\begin{equation}
r_A = 2.6\times10^8   B^{4/7} r_{ns}^{10/7} M_{ns}^{1/7} L_x^{-2/7},
\end{equation}

\noindent
where B is the magnetic field in units of 10$^{12}$ G, r$_{ns}$ the 
neutron star radius in units of 10 km, M$_{ns}$ the neutron star
mass in units of \Msun, and the X-ray luminosity L$_x$ in units of 
10$^{37}$ \ergsec. Figure~\ref{figure8} shows the evaluation of 
Equ. 4 and 5 for two different neutron star radii, 10 km (black) representing the 
lower end of most equations of state, and 15 km (red) representing the
higher end of most equations of state~\citep{oezel16}. Equation 5 has only a 
weak dependence on neutron star mass and here we used a canonical value
of 1.4 \Msun. For L$_x$ we applied the measured X-ray luminosity
in Table~2. For a distance of 9.4 kpc~\citep{heinz15} we can determine
limits to the blackbody emission radius depending on the blackbody fits 
provided in Table~2. This then amounts to  
a r$_{bb}$=r$_{min}$=0.512 km and a r$_{bb}$=r$_{max}$=0.597 km. 
At a distance of 9.4 kpc the observed emission radius 
in Table~4 is 0.589$\pm$0.151 km consistent with these values.
A relativistic color correction would do little to the range of these 
values.  From that we find a lower limit to the magnetic field strength
of 4.0$\times10^{10}$ G for a 15 km radius and and 2.8$\times10^{11}$ G
for a 10 km radius. Note that these are only lower limits and the 
actual field can still be higher then these values.  

\vspace{0.3cm}
\includegraphics[angle=0,width=8.5cm]{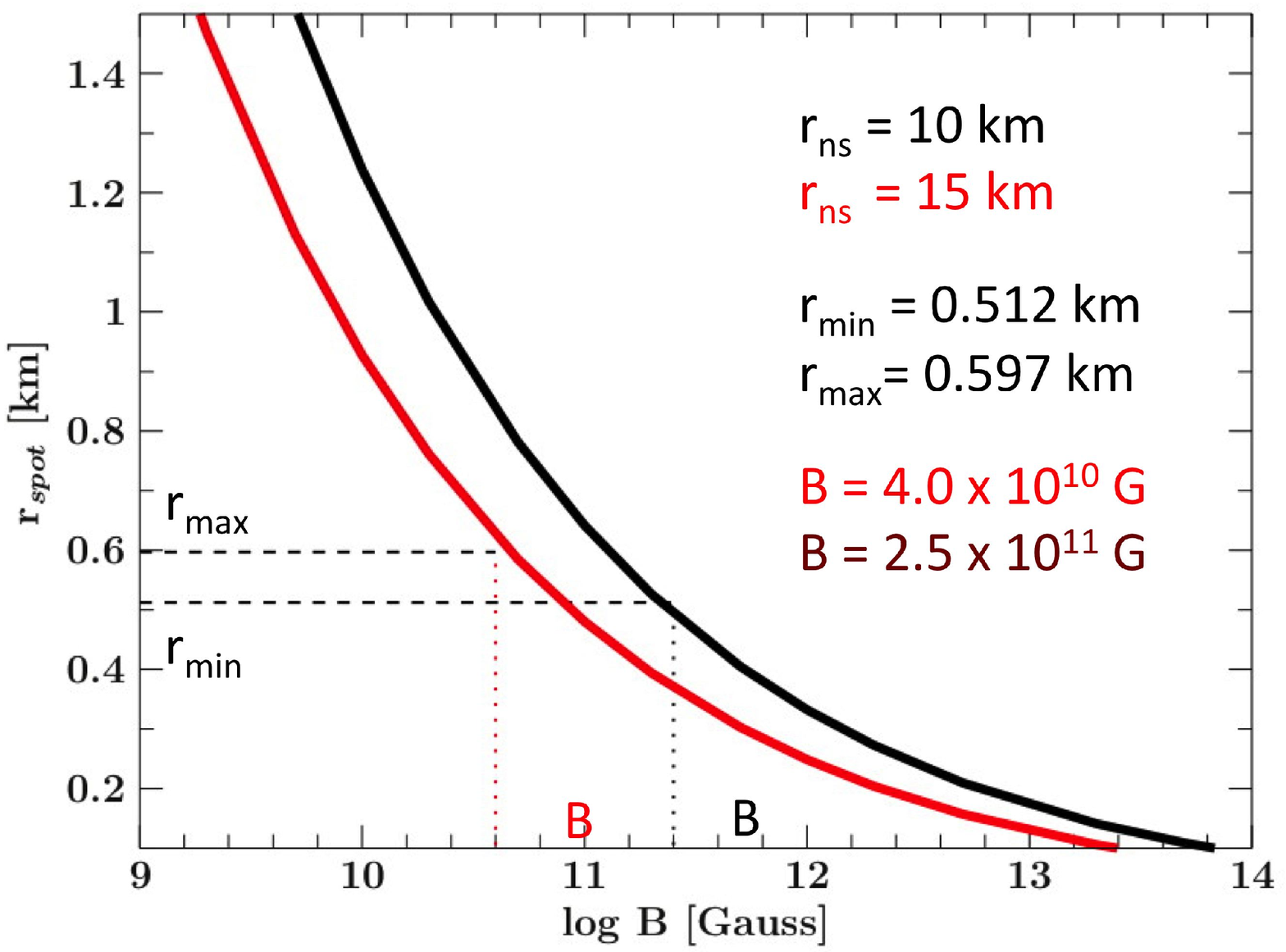}
\figcaption{Evaluations of Equ. 4 and 5 for a neutron star radius of
10 km (black) and 15 km (red). The limits r$_{min}$ and r$_{max}$ are derived from
the blackbody fit results in Table~2 for a distance of 9.4 kpc. 
\label{figure8}}
\vspace{0.3cm}

A survey of about 20 confirmed cyclotron line energies in HMXBs~\citep{caballero12} shows 
that all but one magnetic surface field strengths are above 10$^{12}$ G, except
for Swift J1626.6-5156~\citep{decesar09} which has a field slightly below 10$^{12}$ G.
The result in Figure~\ref{figure8} of course depends on distance, while a cyclotron line 
measurement does not~\citep{coburn06}. 
However, we deem a distance of 9.4 kpc quite robust. In order to have both
the 10 km and 15 km solutions above 10$^{12}$ G, the distance has to be below 6 kpc,
which is implausible because of the amount of extinction along the line of sight and
it would take the X-ray luminosity in observations I and II~\citep{brandt00, schulz02}
too far away from Eddington conditions to produce such a powerful disk wind.
Higher distances do not change the result much unless they become unrealistically
high. In order for the field values to drop below 10$^{10}$ G, the blackbody radius
has to be higher than 1 km and the source would
reside outside the Milky Way. We therefore are confident that the surface magnetic field
in Cir X-1 is somewhere between the values shown in Fig.~\ref{figure8}. 

The moderately high $>> 10^{10}$ G magnetic field
should still be consistent with a young system even though it is generally
assumed that neutron stars are born with $> 10^{12}$ G (see \citet{kaspi10} and
references therein). The neutron star in Kes 79~\citep{halpern10} with a
magnetic field of only $~ 3.1\times10^{10}$ G is a good example
for a younger neutron star with a moderate 
magnetic field (see also \citet{shabaltas12}). Other examples are 
PSR J1852+0040 and 1E1207.4-5209~\citet{halpern07, gotthelf07} as well
as PSR J0821-4300 in Puppis A \citep{gotthelf09}, even though these may not be
quite as young as Cir X-1. Thus young neutron stars
with lower magnetic fields of the order of $10^{11}$ G, 
also sometimes dubbed `anti-magnetars', are not
unusual anymore.

The observation of type I X-ray thermonuclear
bursts~\citep{tennant86, linares10} also implies a field of $<< 10^{12}$ G~\citep{fujimoto81}.
However, this criteria is more based on the lack of observations of type I bursts
in X-ray pulsars rather than having a solid theoretical basis other than the suppression
of convective motions to allow runaway burning in magnetic fields of 
about 10$^{12}$ G and higher~\citep{gough66, bildsten95}.
\citet{bildsten98} outlined the conditions for neutron star nuclear burning
for the case of high mass accretion rates (\mdot $> 10^{10}$ \Msun yr$^{-1}$). At these
rates unstable nuclear burning is more easily realized. The luminosity of Cir X-1 during
its outburst in 2010 was still below the stable burning criterion and here the
magnetic field may have been high enough to confine the accreted matter to the ignition
pressure, but still low enough to allow convective motions. We therefore argue that
in the case that the neutron star magnetic field in Cir X-1 is only moderately high, the
two facts that the neutron is very young and that we observe type I X-ray thermonuclear bursts
are consistent with what we know today.

\subsection{The Nature of the Blueshifts}

The most prominent difference in the observed line properties besides the
fact that the He-like resonance lines appear enhanced are the blueshifts 
of about of 400 \kms in most of the observed lines. In \citet{schulz08}
the lines were at rest and identified as ADC emissions from the accretion disk.
In this low state blueshifted line emissions now need a two orders of magnitude
higher ionizing luminosity to sustain such an ADC in the accretion disk.
The flux of the lines and the amount of blue-shift in the lines appear small, and 
may not have significantly affected the line profiles in observations III and IV 
spcifically  because the shifts are of the order of a grating spectral resolution 
element. However we note that \citet{iaria08} claim a faint blue component
in their line profile analysis in observation III. This could mean that
this blue-shifted emission is always present but gets overpowered by ADC emissions
when the source gets brighter.

There is no viable explanation for blue-shifted lines within 
the accretion disk suggesting emission regions outside the disk. While
\citet{iaria08} proposed possible jet emissions, the determined ionization
parameters as well as emission measures and the amount of the blueshift
are also consistent with X-ray illumination of a massive companion wind. 
A wind velocity of 400 \kms would point 
in direction of a B5Ia supergiant~\citep{prinja98} as a companion confirming
the identification from ~\citet{jonker07}. There is not much room for later types as in this
spectral type range terminal wind velocities decline rapidly with spectral type.
We point out that B5 stellar winds themselves cannot produce X-rays through 
inner wind shocks as we observe in more massive stars. In this case the very low density
outer terminal velocity zone of the wind gets illuminated by the X-ray source
in Cir X-1. 
That we may observe the illumination of the stellar wind is also 
supported by an apastron observation
taken during a more prominent outburst in 2010. Figure~\ref{figure7} shows a dominant
continuum but no photo-ionized lines, which should have been detectable. Even with the 
brighter continuum the three brightest lines in the periastron spectrum should still be  
detectable above a 5$\sigma$ level. Yet the only line detected is the Fe K 
fluorescence line which we commonly associate with the accretion disk.
The absence of unshifted ADC lines at least with regards to S, Si and Mg
from the accretion disk is peculiar because if it is true that the X-ray
emissions have not changed in luminosity and are further obscured than
what we observed in observations III and IV, then these lines should still
be there unless they are now also suppressed by heavy absorption. In that
case the photoionized emissions from the stellar wind is all thats left
to the observer as well as some weak higher Z lines from the ADC. 

The amount of the shift would also rule out much later types as terminal wind velocites decline
rapidly with type~\citep{lamers95, prinja98}. Observations V and VIIa-c were taken
between orbital phase 0.00 and 0.07 using the RXTE/ASM ephemeris (MJD = 50082.04) and
orbital period of 16.54694 days from ~\citet{shirey98} (see also ~\citet{clarkson04})
for which the neutron star always illuminates
the face of the star into a wind coming towards the observer, hence the blueshift.
However, we point out that 400 \kms is also near the known rotational velocities 
of any Be-star in an X-ray binary~\citep{reig11}. So if it is not the wind but
the stellar surface that gets ionized then we should see blue- and red-shifts depending 
on whether the compact object comes towards the companion or moves away and none
when the compact object is a zero phase.
With all the distractions from disk emissions in the previous observations gone, we believe
that is the first time we might see the companion wind itself. 
 

For this one can estimate the contribution of a spherical B-star wind with a constant
velocity and a specific mass loss rate to the emission measure by integrating over the 
available volume from a B-star radius of 10$^{12}$ cm to infinity. With a spherical mass loss
rate of 10$^{-8}$ \Msun yr$^{-1}$ and a terminal wind velocity of 500 \kms this results
to emission measures of the order of 10$^{55}$ erg cm$^{-3}$ s$^{-1}$, which is only
a fraction of what we observe. Increased emission measures require increased
mass loss rates as well as non-spherical geometries. This indicates that if this is 
indeed the companion wind, it has to be enhanced. One indicator that this is the case could be the
existence of significant amounts of resonance scattering in the He-like lines.

\subsection{On the Possible High Mass Nature of the X-ray Binary}

The photoionized plasma
requires a substantial absorber in the line of sight in addition to
absorption from the ISM. If the wind is indeed the emission line region, 
this absorber has to be able to cover the bulk of the 
stellar wind. Normal B5 supergiants are not well known to 
carry wind produced stellar disks. The only plausible protagonists we know today are 
magnetic stars and Be-stars, in the latter case also mostly much earlier type B-stars. 
In both cases an equatorial wind fed decretion
disk can be produced under various critical circumstances. 
In the case of Be stars it may be a stellar rotation rate of 
more than 75$\%$ of the critical rate (see \citet{rivinius13} for a full review),
while in the case of magnetic stars it is a strong magnetic field configuration confining
the wind (see \citet{gagne05}). Such disks can provide not only the additional line 
of sight absorber but also account for the resonance scattering we observe. 

However, while this is still a lot of speculation, we can in the following consider 
what is observationally known for the case of Be stars. 
Be-star X-ray binaries (BeXBs) are a sizeable sub-group in the category of high-mass X-ray 
binaries (HMXBs). Perhaps the most famous in the class of BeXBs is
GX 301-2. This system consists of an accreting magnetized neutron star in an 
eccentric orbit (e$\sim$0.46; \citet{sato86, koh97}). This is a very similar
system except its orbit is three times as long and periastron passages
are not as violent as in Cir X-1. Its optical counterpart is a B1.5Ia supergiant
with a lower mass limit of 39 \Msun,
which is much larger than what is considered for Cir X-1 and here the Roche lobe overflow connection
never breaks~\citep{sato86, watanabe03, kaper06}. However, there are many similarities,
such as an inclination larger than 44$^{o}$ which does not produce an eclipse but
just barely misses the face of the star producing dips in the light curve by the
dense stellar wind. \citet{jonker07} also concluded even though the inclination is high
\citep{brandt96} that in the case of Cir X-1 the neutron star never crosses
the star (see also \citet{iaria08}). The X-ray spectrum in observation III~\citep{schulz08}
is very similar to the one observed in GX 301-2, where the region above 5~\AA\ is dominated
by line emissions from the ionized wind and below is dominated by accretion disk activity
~\citep{sato86, koh97, watanabe03, fuerst11}.

In Be stars the regions above and below the disk are more or less equivalent to those
surrounding normal B-stars~\citep{rivinius13}. Fast rotation may even enhance the wind
towards the polar regions~\citep{puls08}. The fact that the companion in Cir X-1 is 
a supergiant does not make much of a difference. Most massive stars begin the main
sequence as fast rotators and usually stay that way~\citep{langer97}, during the 
supergiant phase the star may see a reduction in the rotation rate due to the increasing
stellar radius~\citep{puls08}. Generally there is no reason why a fast rotator would
not retain its disk in the supergiant phase and the case could be made that
Cir X-1 might as well be the youngest known BeXB.  

All BeXBs classified so far have orbital periods of 20 days and higher up to 300 days
(see ~\citet{reig11} for a review) and Cir X-1 with 16.5 days would mark the
shortest period, which seems adequate since it would be the youngest in the sample.
However, the list of known OBe stars mostly shows dwarfs (class V), subgiants (class IV)
or giants (class III) and Cir X-1 would be the only other supergiant case (classes I and II)
to GX 301-2.
Supergiant X-ray binaries (SGXBs) tend to have shorter periods and here Cir X-1 
would have the longest of this class. It is also the case that almost all BeXBs show
X-ray pulsations ranging from a few seconds to a few hundreds of seconds. To date,
no pulsations have been reported for Cir X-1 and we also did not detect any in the range
where we are sensitive which is also above a few seconds. For the neutron star being
that young we should also expect much shorter periods likely down to a few milliseconds.
The lower magnetic field of 10$^{11}$ G should not brake its period as fast as the
other cases in 4000 yr. The mere fact that we do not observe pulsations is 
somewhat of concern but not completely unusual. To date no pulsations have been 
found in the SGXB 4U1700-37~\citep{seifina16} and in a few BeXBs~\citep{reig11}.
New model calculations also show that at moderate magnetic field strength of the 
order of 10$^{11}$ G and a high inclination of the system~\citep{brandt95},
pulsations are quite impossible 
to detect if the angle between the neutron star rotation and magnetic field axis
is small (Falkner et al. 2019, submitted
\footnote{https://www.sternwarte.uni-erlangen.de/docs/theses/2018-07$\_$Falkner.pdf}).
We should also mention that \citet{cumming08} proposed for the case of HETEJ1900.1-2455
and possibly other accreting millisecond pulsars
that B-fields can be partially buried suppressing the observation of pulsations.

The fact that we know that the neutron star in Cir X-1 is very young combined with
an identification of the companion as a B5Ia supergiant also can put some constraints
on the evolutionary state of the entire binary.  
A formation scenario involving an AIC~\citep{bhattacharya91} now seems
highly unlikely.
It also rules out that the
progenitor star was a massive O-star as these only live less than 20 Myr whereas
B3 stars and later live more than 30 Myr~\citep{behrend01}. Since the companion
in Cir X-1 is in its supergiant phase, which the star is in during its very late
main sequence state, it must be much older than 20 Myrs. More plausible is that
the progenitor was quite similar to the companion, maybe one or two types earlier.
This would put the progenitor into likely the lowest mass range (8 - 10 \Msun) that
can produce a neutron star. Today we know little about neutron star progenitor
masses and if this conclusion is true, it is quite extraordinary.

Last but not least the classification of Cir X-1 as a BeXB might have the 
potential to at least partially explain the $\sim$30 yr transient flux
behavior as shown by \citet{parkinson03} by invoking a precession period for the 
Be-star-disk system. Such precession scenarios have already been suggested by 
\citet{brandt95} in terms of accretion disk precession and by \citet{heinz13}
in terms of spin-orbit coupling effects between the neutron star spin and 
the binary orbit. Here we suggest a precession of a companion star disk.
Super-orbital periods in accretion disks are not unusual
in X-ray binaries as the examples of Her X-1, LMC X-4 and SMC X-1 show. 
Precessing Be-star disks are much more rare but not unheard of.  
\citet{lau16} recently reported on an apparent precessing helical
outflow from the massive star WR102c. The interpreation is that the precessing outflow
emerged from a previous evolutionary rapidly rotating phase of the star and 
attributed the precession to an unseen compact companion. In a sense this situation
is not unsimilar to what we envision here. In the WR102c system the period 
of the unseen companion was constrained to between 800 and 1400 days, which is
much larger than the period in Cir X-1, but the precession period of 
1.4$\times10^4$ yr is very close to the long term variation cycle in Cir X-1.
This sets a precedent and it is no longer a question whether in can happen,
but it calls for details of how it happens. 

\acknowledgments

\bibliographystyle{jwapjbib}
\bibliography{mnemonic,jw_abbrv,apj_abbrv,ms}
\end{document}